\documentstyle[twoside,psfig,fleqn,espcrc2]{article}
\newcommand {\ignore}[1]{}

\newcommand{\bc}{\begin{center}}
\newcommand{\ec}{\end{center}}

\def\ifmath#1{\relax\ifmmode #1\else $#1$\fi}

\def\3quarter{{\textstyle{3 \over 4}}}

\def\ra{\rightarrow}

\def\lf{\leaders\hbox to 1em{\hss.\hss}\hfill}
\def\21{$SU(2) \ot U(1)$}
\def\321{$SU(3) \ot SU(2) \ot U(1)$}
\def\ne{\hbox{$\nu_e$ }}
\def\nm{\hbox{$\nu_\mu$ }}
\def\nt{\hbox{$\nu_\tau$ }}
\def\ns{\hbox{$\nu_{s}$ }}

\def\neus{\hbox{neutrinos }}

\def\neu{\hbox{neutrino }}

\def\eq#1{{eq. (\ref{#1})}}

\def\fig#1{{Fig. \ref{#1}}}

\def\VEV#1{\left\langle #1\right\rangle}
\let\vev\VEV

\def\ltap{\raisebox{-.4ex}{\rlap{$\sim$}} \raisebox{.4ex}{$<$}}

\def\lsim{\raise0.3ex\hbox{$\;<$\kern-0.75em\raise-1.1ex\hbox{$\sim\;$}}}
\def\gsim{\raise0.3ex\hbox{$\;>$\kern-0.75em\raise-1.1ex\hbox{$\sim\;$}}}

\def\beq{\begin{equation}}
\def\eeq{\end{equation}}
\def\bef{\begin{figure}}
\def\eef{\end{figure}}
\def\bet{\begin{table}}
\def\eet{\end{table}}
\def\bea{\begin{eqnarray}}
\def\ba{\begin{array}}
\def\ea{\end{array}}
\def\bi{\begin{itemize}}
\def\ei{\end{itemize}}
\def\ben{\begin{enumerate}}
\def\een{\end{enumerate}}
\def\ra{\rightarrow}
\def\ot{\otimes}
\def\eea{\end{eqnarray}}
\def\apj#1#2#3{          {\it Astrophys. J. }{\bf #1} (19#2) #3}
\def\app#1#2#3{          {\it Astropart. Phys. }{\bf #1} (19#2) #3}

\def\ib#1#2#3{           {\it ibid. }{\bf #1} (19#2) #3}
\def\nat#1#2#3{          {\it Nature }{\bf #1} (19#2) #3}
\def\nps#1#2#3{        {\it Nucl. Phys. B (Proc. Suppl.) }{\bf #1} (19#2) #3} 
\def\np#1#2#3{           {\it Nucl. Phys. }{\bf #1} (19#2) #3}
\def\pl#1#2#3{           {\it Phys. Lett. }{\bf #1} (19#2) #3}
\def\pr#1#2#3{           {\it Phys. Rev. }{\bf #1} (19#2) #3}
\def\prep#1#2#3{         {\it Phys. Rep. }{\bf #1} (19#2) #3}
\def\prl#1#2#3{          {\it Phys. Rev. Lett. }{\bf #1} (19#2) #3}

\def\rmp#1#2#3{          {\it Rev. Mod. Phys. }{\bf #1} (19#2) #3}
\def\zp#1#2#3{           {\it Zeit. fur Physik }{\bf #1} (19#2) #3}

\def\n.c.#1#2#3{         {\it Nuovo Cim. }{\bf #1} (19#2) #3}
\def\r.n.c.#1#2#3{       {\it Riv. del Nuovo Cim. }{\bf #1} (19#2) #3}
\def\sjnp#1#2#3{         {\it Sov. J. Nucl. Phys. }{\bf #1} (19#2) #3}

\def\jetp#1#2#3{         {\it JETP }{\bf #1} (19#2) #3}
\def\mpl#1#2#3{          {\it Mod. Phys. Lett. }{\bf #1} (19#2) #3}

\def\ppnp#1#2#3{           {\it Prog. Part. Nucl. Phys. }{\bf #1} (19#2) #3}

\def\tp{these proceedings}

\def\bnue{\hbox{$\bar\nu_e$ }}

\newcommand{\AmS}{{\protect\the\textfont2
 A\kern-.1667em\lower.5ex\hbox{M}\kern-.125emS}}
\title{Neutrinos and Physics Beyond the Standard Model}
\author{J. W. F. Valle\address{Instituto de F\'{\i}sica Corpuscular 
- C.S.I.C.\\Departament de F\'{\i}sica Te\`orica, Universitat de 
Val\`encia\\46100 Burjassot, Val\`encia, Spain
}
\thanks{ Supported by DGICYT grant  PB95-1077 and in part by EEC 
under the TMR contract ERBFMRX-CT96-0090. I thank the organizers
for the hospitality, and H. Nunokawa and F. Vissani for helping
me prepare the manuscript and M. A. Diaz for proof-reading it.}}
\begin{document}

\hfill hep-ph/9702231\\
\hfil FTUV/97-5 \\
\hfil IFIC/97-5 

\begin{center}
{\Large Neutrinos and Physics Beyond the Standard Model}
\end{center}

\begin{center}
J. W. F. Valle\\
Instituto de F\'{\i}sica Corpuscular - C.S.I.C.\\
Departament de F\'{\i}sica Te\`orica, Universitat de Val\`encia\\
46100 Burjassot, Val\`encia, Spain
\end{center}

A brief sketch is made of the present observational status of 
neutrino physics, with emphasis on the hints that follow from 
solar and atmospheric neutrino observations, as well as 
cosmological data on the amplitude of primordial density 
fluctuations. I also briefly review the ways to account 
for the observed anomalies and some of their implications.

\begin{abstract}
A brief sketch is made of the present observational status of 
neutrino physics, with emphasis on the hints that follow from 
solar and atmospheric neutrino observations, as well as 
cosmological data on the amplitude of primordial density 
fluctuations. I also briefly review the ways to account 
for the observed anomalies and some of their implications.
\end{abstract}
\maketitle

%

%

\section{INTRODUCTION}
\vskip .3cm

Two puzzles exist associated to neutrinos: if massless, 
they would be only fermions with this property. If massive,
why are their masses so much smaller than those of their
charged brothers? The fact that neutrinos are the only
electrically neutral elementary fermions may hold the
key to the answer, namely neutrinos could be Majorana 
fermions, in some sense the most fundamental ones. In
this case the supression of their mass could be associated 
to lepton number conservation. But so far we do not know.

It is beyond any doubt that, one of the most fundamental 
drawbacks of the Standard Model is the fact that it says 
so little about the properties of neutrinos. Their 
masslessness is not dictated by a fundamental underlying 
principle, such as gauge invariance in the case of the photon. 
Most extensions of the Standard Model require neutrinos
to be massive. One interesting aspect of many models
where neutrinos have non-vanishing masses is that they 
lead to effects that could be experimentally tested, even
outside the conventional realm of neutrino experiments,
such as underground experiments. In some cases one may 
probe, though indirectly, the underlying physics at 
high energy accelerators.

From the observational point of view nonzero 
neutrino masses now seem required in order to account 
for the data on solar and atmospheric neutrinos, as well as 
the (hot) dark matter in the universe \cite{warsaw}. But before 
overviewing the present observational limits and hints in favour 
of massive neutrinos, let us make a few general remarks about the 
theoretical models \cite{fae}.

\section{MODELS}
\vskip .3cm

One of the simplest extensions of the electroweak theory 
consists in adding isosinglet neutral heavy leptons (NHLS), 
such as right handed neutrinos, as in the seesaw model \cite{GRS}. 
In this case the NHLS have a large Majorana mass term $M_R$, which 
violates total lepton number, or B-L (baryon minus lepton number), a 
symmetry that plays an important role in many extended gauge 
models \cite{LR}. The masses of the light neutrinos are
obtained by diagonalizing the following mass matrix
\beq
\begin{array}{c|cccccccc}
& \nu & \nu^c\\
\hline
\nu   & 0 & D \\
\nu^c & D^T & M_R
\end{array}
\label{SS}
\eeq
where $D = h_D \vev{H} /\sqrt2$ is the Dirac mass matrix and
$M_R = M_R^T$ is the isosinglet Majorana mass. In the seesaw 
approximation, one finds 
\beq
{M_L} = - D M_R^{-1} D^T \:.
\label{SEESAW}
\eeq
This mechanism is able to explain naturally the relative smallness of 
\neu masses \cite{GRS}. Although the seesaw idea was suggested in the 
context of SO(10) or left-right symmetric extensions where lepton 
number is a part of the gauge symmetry \cite{LR}, it may be directly 
introduced in the Standard Model. Though it is natural to expect $M_R$ 
to be large, one can not make any firm guess, as its magnitude heavily 
depends on the model. As a result one can not make any real 
prediction for the corresponding light neutrino masses that are 
generated through this so-called  {\sl seesaw mechanism} \cite{GRS}. 

Although attractive, the seesaw mechanism is by no means 
the only way to generate neutrino masses. There is a large 
diversity of possible schemes to generate neutrino masses, 
which do not require any new large mass scale. For example, 
it is possible to start from an extension of the lepton sector 
of the \21 theory by adding a set of $two$ 2-component isosinglet 
neutral fermions, denoted ${\nu^c}_i$ and $S_i$. In this case
there is an exact L symmetry that keeps neutrinos strictly massless, 
as in the Standard Model. The conservation of total lepton number 
leads to the following form for the neutral mass matrix
\beq
\begin{array}{c|cccccccc}
& \nu  & \nu^c & S \\
\hline
\nu & 0 & D & 0 \\
\nu^c & D^T & 0 & M \\
S & 0 & M^T & 0
\end{array}
\label{MAT}
\eeq
This form has also been suggested in various theoretical models
\cite{WYLER}, including many of the superstring inspired models. 
In the latter case the zeros of \eq{MAT} naturally arise 
due to the absence of Higgs fields to provide the usual Majorana 
mass terms, needed in the seesaw model \cite{SST}. The implications
of \eq{MAT} are interesting on their own right, and the model 
represents a conceptually simple and phenomenologically rich
extension of the Standard Model, which brings in the possibility
that a wide range of new phenomena be sizeable. These have to do 
with neutrino mixing, universality, flavour and CP violation in 
the lepton sector \cite{BER,CP}, as well as direct effects 
associated with the NHL production in $Z$ decays \cite{CERN}.
A remarkable feature of this model is the possibility of 
non-trivial neutrino mixing  despite the fact that neutrinos 
are strictly massless. This tree-level effect was exploited in 
ref. \cite{massless0,massless}. Moreover, there are loop-induced
lepton flavour and CP non-conservation effects whose rates are 
precisely calculable  \cite{BER,CP,3E}. I repeat that this is
remarkable due to the fact that physical light neutrinos are 
massless, as in the Standard Model. This feature is the same 
as what happens in the supersymmetric mechanism of flavour 
violation \cite{Hall}. Indeed, in the simplest case of SU(5) 
supergravity unification, there are flavour violating processes, 
like $\mu \ra e \gamma$, despite the fact that in SU(5) neutrinos 
are protected by B-L and remain massless. The supersymmetric
mechanism and that of \eq{MAT} differ in that the lepton flavour 
violating (LFV) processes are induced in one case by NHL loops, 
while in supersymmetry they are induced by scalar boson loops. 
In both cases the particles in the loops have masses at the 
weak scale, leading to potentially sizeable rates. The allowed 
LFV branching ratios in supersymmetry have been  discussed 
here by Barbieri \cite{SUSYLFV,SUSYLFV2}. In the NHL model,
as a result of the masslessness of the physical neutrinos, 
the allowed value for the LFV branching ratios is also 
maximized in the class of neutrino mixing models. Indeed,
one can show that some of the LFV rates can be sizeable 
enough to be experimentally detectable \cite{ETAU,TTTAU,opallfv}.  

 Clearly, one can easily introduce non-zero masses in this
model through a $\mu S S$ term that could be proportional
to the VEV of a singlet field $\sigma$ \cite{CON}. In contrast 
to the seesaw scheme, the \neu masses are directly proportional to
$\VEV{\sigma}$, a fact which is very important for the phenomenology
of the Higgs boson sector.

\begin{figure}
\centerline{
\psfig{file=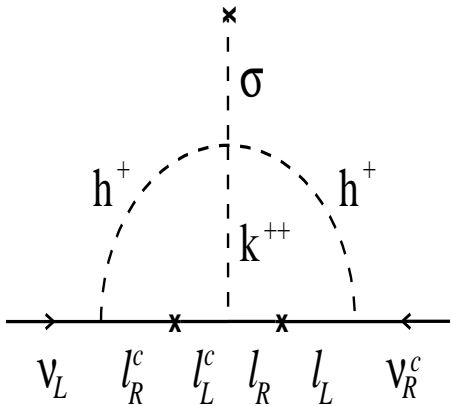,height=5.4cm,width=6cm}}
\caption{Two-loop-induced Neutrino Mass. }
\vglue -.5cm
\label{2loop}
\end{figure}

There is also a large variety of possible {\sl radiative} schemes 
to generate \neu masses. The prototype models of this type are the
Zee model and the model suggested by Babu \cite{zee.Babu88}. In these
models lepton number is explicitly broken, but it is easy to realize
them with spontaneous breaking of lepton number. For example in 
the version suggested in ref. \cite{ewbaryo} the neutrino mass arises
from the diagram shown in \fig{2loop}
These models  do not require one to introduce a large 
mass scale. It is quite possible to embed such schemes so as to have 
the spontaneous violation of the global lepton number symmetry. 
In these models again the \neu masses are directly 
proportional to $\VEV{\sigma}$ or some positive power of that,
depending on the lepton numbers assignments of the model. Thus
the scale at which such a symmetry gets broken does not need to 
be high, as in the original proposal \cite{CMP}, but can be rather 
low, close to the weak scale \cite{JoshipuraValle92}. Such models 
are very attractive and lead to a richer phenomenology, as the 
extra particles required have masses at scales that could be 
accessible to present experiments.

\section{LIMITS ON NEUTRINO MASSES AND MIXINGS}
\vskip .3cm

\subsection{Laboratory Limits }
\vskip .3cm

The best limits on the neutrino masses can be summarized as
\cite{PDG96}:
\beq
\label{1}
m_{\nu_e} 	\lsim 5 \: \rm{eV}, \:\:\:\:\:
m_{\nu_\mu}	\lsim 170 \: \rm{keV}, \:\:\:\:\:
m_{\nu_\tau}	\lsim 23 \: \rm{MeV}
\eeq
These are the most model-independent of the laboratory limits 
on \neu mass, as they follow purely from kinematics.
The limits on the \ne mass comes from beta decay, that on the 
\nm mass comes from PSI (90 \% C.L.) \cite{psi}, with further 
improvement limited by the uncertainty in the $\pi^-$ mass. 
On the other hand, the best \nt mass limit now comes from high 
energy LEP experiment ALEPH \cite{eps95} and may be substantially 
improved at a future tau-charm factory \cite{jj}. In  connection 
with tritium beta decay limit \cite{Lobashev} it is interesting 
to remark  two features in the spectrum. One is an excess of 
events near the end point of the differential spectrum, at 
$Q - E_e \lsim 10$ eV, leading to a negative $m^2$ value in 
the fit and probably of instrumental origin. The second is
an excess of events at lower electron energies 
$Q - E_e \gsim 200$ eV, also observed by the Mainz group.  
One possible explanation of this anomaly is the existence 
of a neutrino  with mass $m \sim 200$ eV and mixing with
the \ne with a probability $P \sim 1 - 2$ \% . It has been
noted that this is precisely in the range implied by 
pulsar velocities \cite{warsaw}.

\begin{figure}
\psfig{file=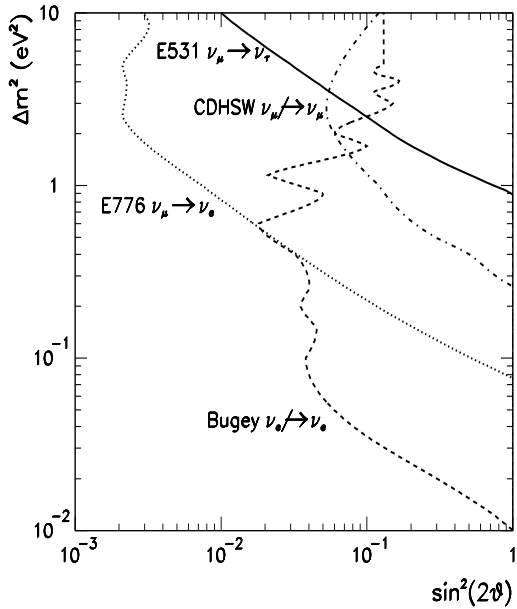,height=8.5cm,width=7cm}
\vglue -0.5cm
\caption{Limits on oscillation parameters.}
\vglue -0.5cm
\label{oscil}
\end{figure}

In addition, there are limits on neutrino masses that follow 
from the non-observation of neutrino oscillations. 
The most stringent bounds come from reactor oscillation experiments 
\cite{reactor} (\bnue - $\nu_x$ oscillations); from
meson factory oscillation experiments (KARMEN \cite{karmen}, 
LSND \cite{lsnd}) and from accelerator experiments E531 
(\nm - \nt) \cite{E531} and E776 (\nm - \nt) \cite{E776}.
The 90 \% confidence level (C.L.) exclusion contours of neutrino 
oscillation parameters in the 2-flavour approximation are 
summarized in Fig. \ref{oscil}, taken from ref. \cite{jjc}. 
Improvements are expected from the ongoing CHORUS and NOMAD 
experiments at CERN, with a similar proposal at Fermilab 
\cite{chorus}. As we have heard at this conference, there 
are also good prospects for substantial progress at future 
long-baseline experiments using KEK, CERN and Fermilab 
neutrino beams aimed at the Kamiokande, Gran Sasso and 
Soudan underground facilities, respectively. For recent
theoretical studies on the possibility of performing
CP violation studies in long-baseline neutrino oscillation 
experiments see ref. \cite{cplong}.

If neutrinos are of Majorana type a new form of nuclear double 
beta decay would take place in which no neutrinos are emitted in
the final state, i.e. the process by which an $(A,Z-2)$ nucleus 
decays to $(A,Z) + 2 \ e^-$. In such process one would have a 
virtual exchange of Majorana neutrinos. Unlike the ordinary double 
beta decay process, the neutrino-less process violates lepton number 
and its existence would indicate the Majorana nature of neutrinos.
Because of the phase space advantage, this process is a very 
sensitive tool to probe into the nature of neutrinos.
Present data place an important limit on a weighted average \neu 
mass parameter $\VEV{m}$ \cite{Klapdor}
\beq
\label{bb}
\VEV{m} \lsim 1 - 2 \ eV
\eeq
depending to some extent on the relevant nuclear matrix elements 
characterising this process \cite{haxtongranada}. This bound
comes from the Heidelberg-Moscow experiment and there might be 
further improvements, also from the IGEX experiment.
Note that the parameter $\VEV{m}$ involves both neutrino masses 
and mixings. Thus, although rather stringent, this limit in \eq{bb} 
may allow relatively large \neu masses, as there may be strong 
cancellations between different neutrino types. This may happen
automatically in the presence of suitable symmetries. For example,
the decay vanishes if the intermediate neutrinos are Dirac-type, 
as a result of the corresponding lepton number symmetry \cite{QDN}. 

Neutrino-less double beta decay has a great conceptual
importance. It has been shown \cite{BOX} that in a gauge theory
of the weak interactions a non-vanishing ${\beta \beta}_{0\nu}$ 
decay rate requires  \neus to be Majorana particles, 
{\sl irrespective of which mechanism induces it}. This is
important since in a gauge theory neutrino-less  double beta 
decay may be induced in other ways. 

\subsection{Limits from Cosmology }
\vskip .3cm

There are a variety of cosmological arguments that give 
information on neutrino parameters. In what follows I briefly 
consider the critical density and the primordial Nucleosynthesis 
arguments.

\subsubsection{The Cosmological Density Limit }
\vskip .3cm

The oldest cosmological bound on neutrino masses is the
one that follows from avoiding the overabundance of relic 
neutrinos \cite{KT}
\beq
\label{RHO1}
m_{\nu_\tau} \lsim 92 \: \Omega_{\nu} h^2 \: eV\:,
\eeq
where $\Omega_{\nu} h^2 \leq 1$ and the mass is assumed to be
less than $O(1 \: MeV)$. Here $\Omega_{\nu}=\rho_{\nu}/\rho_c$, 
where $\rho_{\nu}$ is the neutrino contribution to the total 
density and $\rho_c$ is the critical density.
The factor $h^2$ measures the uncertainty in the determination of the
present value of the Hubble parameter, $0.4 \leq h \leq 1$. 
The factor $\Omega_{\nu} h^2$ is known to be smaller than 1.

For the $\nu_{\mu}$ and $\nu_{\tau}$ this bound is much more 
stringent than the corresponding laboratory limits \eq{1}. 

Recently there has been a lot of work on the possibility of
an MeV tau neutrino \cite{ma1}. Such range seems to be an 
interesting one from the point of view of structure formation 
\cite{ma1}. Moreover, it is theoretically viable as the
constraint in \eq{RHO1} holds only if \neus are stable on the 
relevant cosmological time scales. In models with spontaneous 
violation of total lepton number \cite{CMP} there are new 
interactions of neutrinos with the majorons which may cause
neutrinos to decay into a lighter \neu plus a majoron, 
for example \cite{fae},
\beq
\label{NUJ}
\nu_\tau \ra \nu_\mu + J \:\: .
\eeq
or have sizeable annihilations to these majorons,
\beq
\label{JJ}
\nu_\tau + \nu_\tau \ra J + J \:\: .
\eeq
The possible existence of fast decay and/or annihilation channels 
could eliminate relic neutrinos and therefore allow them to be heavier
than \eq{RHO1}. The cosmological density constraint on neutrino decay 
lifetime (for neutrinos lighter than 1 MeV or so) may be written as
\beq
\tau \ltap 1.5 \times10^7 (KeV/m_{\nu_\tau})^{2} yr \:,
\label{RHO2}
\eeq
and follows from demanding an adequate red-shift of the heavy neutrino
decay products. For neutrinos heavier than $\sim 1 \: MeV$, such as
possible for the case of $\nu_{\tau}$, the cosmological limit on the
lifetime is less stringent than given in \eq{RHO2}.
 
As we already mentioned the possible existence of non-standard 
interactions of neutrinos due to their couplings to the Majoron
brings in the possibility of fast invisible neutrino decays with 
Majoron emission \cite{fae}. These 2-body decays can be
much faster than the visible decays, such as radiative decays of 
the type $\nu' \ra \nu + \gamma$. As a result the Majoron decays
are almost unconstrained by astrophysics and cosmology. For
a more detailed discussion see ref. \cite{KT}. 

A general method to determine the Majoron emission decay rates 
of neutrinos was first given in ref. \cite{774}. The resulting 
decay rates are rather subtle \cite{774} and model dependent
and will not be discussed here. The reader may consult ref.
\cite{V,fae}. The conclusion is that there are many ways
to make neutrinos sufficiently short-lived that all mass values
consistent with laboratory experiments are cosmologically acceptable.
For neutrino decay lifetime estimates see ref. \cite{fae,V,RPMSW}.

\subsubsection{The Nucleosynthesis Limit}
\vskip .3cm

There are stronger limits on neutrino lifetimes and/or
annihilation cross sections arising from cosmological 
nucleosynthesis considerations. Recent contradictory 
data on the primordial deuterium abundance \cite{dhigh,dlow}
have stimulated a lot of work on the subject \cite{cris,ncris,sarkar}.
 If massive $\nu_\tau$'s are stable on the nucleosynthesis time scale, 
($\nu_\tau$ lifetime longer than $\sim 100$ sec), they can lead to an
excessive amount of primordial helium due to their large contribution 
to the total energy density. This bound can be expressed through 
an effective number of massless neutrino species ($N_\nu$). Using
$N_\nu < 3.4-3.6$, the following range of $\nu_\tau$ mass has been 
ruled out \cite{KTCS91,DI93}
\begin{equation}
0.5~MeV < m_{\nu_\tau} < 35~MeV
\label{cons1}
\end{equation}
If the nucleosynthesis limit is taken less stringent the limit
loosens somewhat. However it has recently been argued that 
non-equilibrium effects from the light neutrinos arising from
the annihilations of the heavy \nt's make the constraint stronger
and forbids all $\nu_\tau$ masses on the few MeV range. 

One can show that if the \nt is unstable during nucleosynthesis 
\cite{unstable} the bound on its mass is substantially weakened 
translated as a function of the assumed lifetime \cite{unstable}.

Even more important is the effect of neutrino annihilations \cite{DPRV}.
Fig. 2 gives the effective number of massless neutrinos equivalent 
to the contribution of massive neutrinos with different values of 
the coupling $g$ between $\nu_\tau$'s and $J$'s, expressed in units 
of $10^{-5}$. For comparison, the dashed line corresponds to the 
Standard Model $g=0$ case. One sees that for a fixed $N_\nu^{max}$, 
a wide range of tau neutrino masses is allowed for large enough 
values of $g$. No \nt masses below 23 MeV can be ruled out, as long 
as $g$ exceeds a few times $10^{-4}$. 
\begin{figure}
\psfig{file=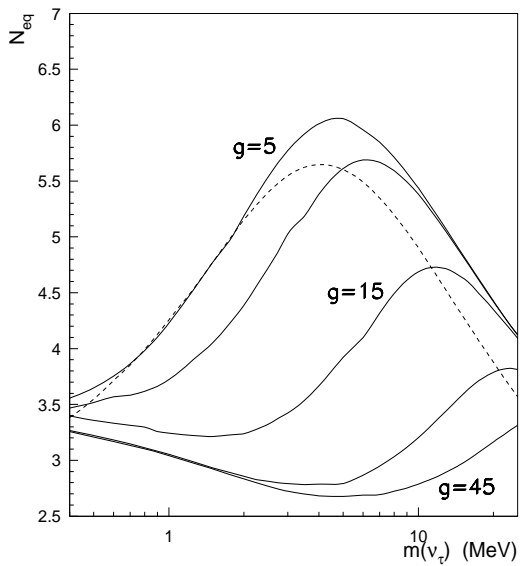,height=7.5cm,width=7cm}
\vglue -.5cm
\caption{Contribution of a heavy \nt to nucleosynthesis
in terms of the equivalent number of massless neutrinos.}
\label{neq}
\end{figure}
One can express the above results in the $m_{\nu_\tau}-g$ plane, 
as shown in  \fig{neffmg}. The region above each curve is 
allowed for the corresponding $N_{eq}^{max}$.  
\begin{figure}
\vglue .5cm
\centerline{
\psfig{file=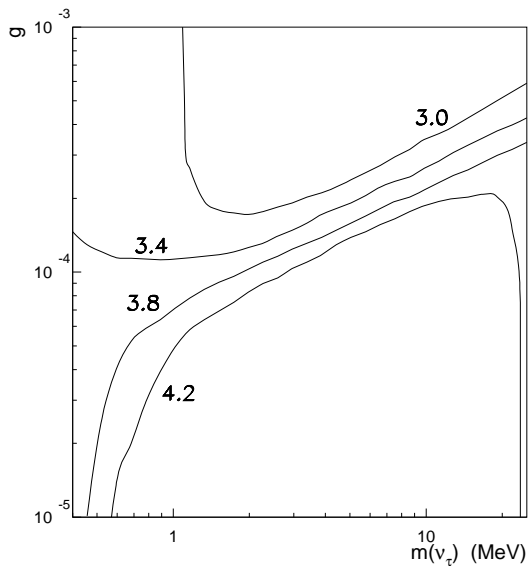,height=7.5cm,width=7cm}}
\vglue -.5cm
\caption{The values of $g(m_{\nu_\tau})$ above each line 
would be allowed by nucleosynthesis if one adopts the $N_{eq}^{max}
= 3, 3.4, 3.8, 4.2$ (from top to bottom). }
\label{neffmg}
\end{figure}
One sees that the constraints on the mass of a Majorana
$\nu_\tau$ from primordial nucleosynthesis can be substantially 
relaxed if annihilations $\nu_\tau \bar{\nu}_\tau \leftrightarrow JJ$ 
are present. It is instructive to notice that the required values of
$g(m_{\nu_\tau})$ are reasonable in many majoron models 
\cite{fae,MASIpot3,DPRV}.

\subsection{Limits from Astrophysics  }
\vskip .3cm

\begin{figure}
\centerline{\protect\hbox{
\psfig{file=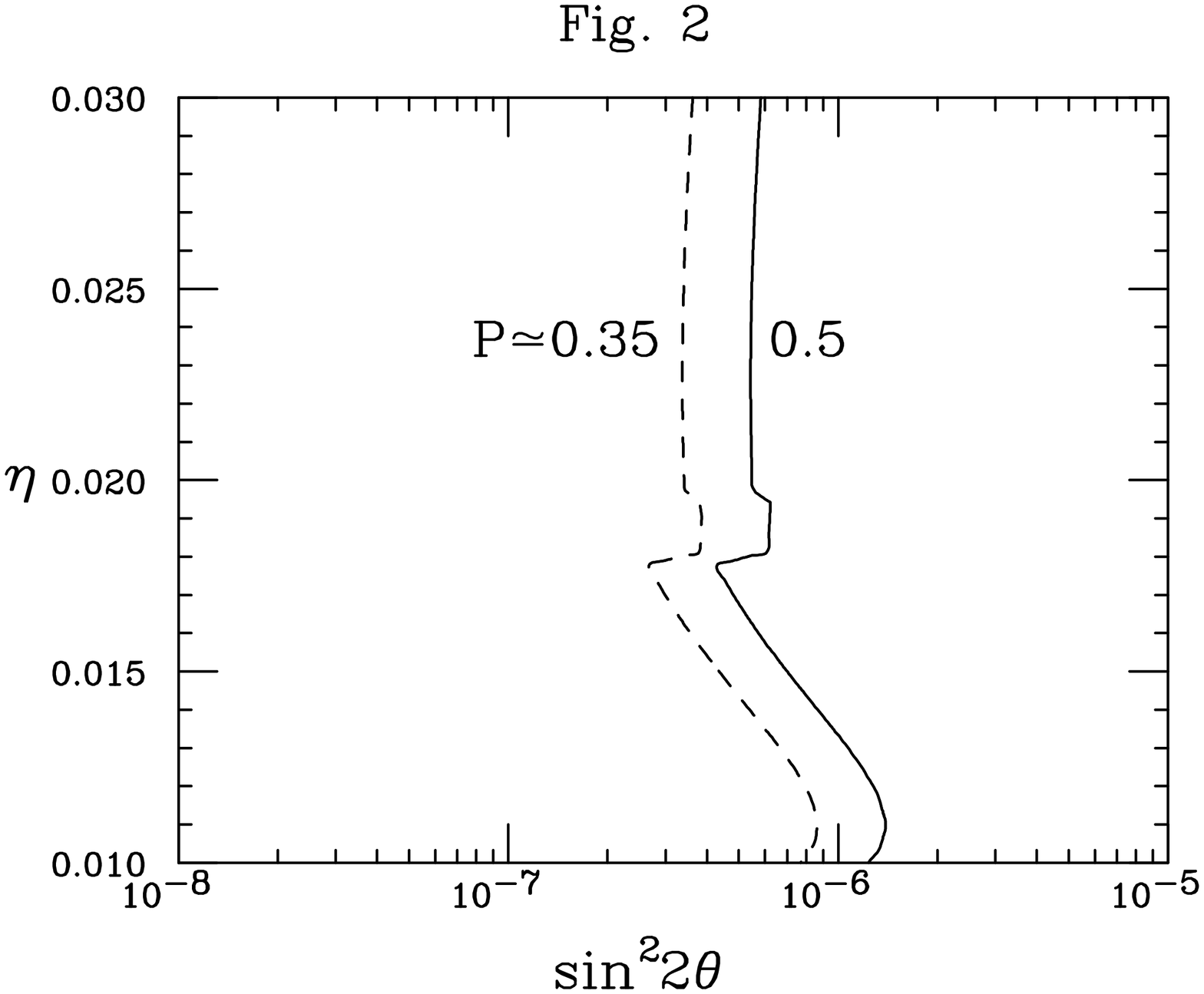,height=7cm,width=8cm,angle=90}
}}
\vglue -7.3cm
\hglue 2.7cm
\psfig{file=cover2.ps,height=0.7cm,width=3cm,angle=90}
\vglue 5.3cm
\caption{
Constraints on massless neutrino mixing from SN1987A. }
\label{SN87}
\vglue -.5cm
\end{figure}

\begin{figure}
\centerline{\protect\hbox{
\psfig{file=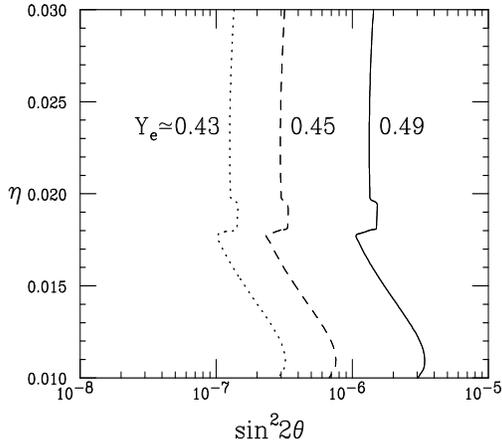,height=7cm,width=8cm,angle=90}
}}
\vglue -7.3cm
\hglue 2.7cm
\psfig{file=cover2.ps,height=0.7cm,width=3cm,angle=90}
\vglue 5.3cm
\caption{Constraints on massless neutrino mixing  arising
 from nucleosynthesis in supernovae.}
\vglue -.5cm
\label{rprocess}
\end{figure}

There are a variety of limits on neutrino parameters
that follow from astrophysics, e.g. from the supernova 1987A 
observations by the Kamiokande and IMB collaborations, as well as from 
the theory of supernovae, including supernova dynamics \cite{Raffelt}
and from nucleosynthesis in supernovae \cite{qian}. Here I briefly
discuss a recent illustration of how supernova physics constrains
neutrino parameters \cite{massless}. 
It has been noted a long time ago that, in some circumstances,
{\sl massless} neutrinos may be {\sl mixed} in the leptonic charged 
current \cite{massless0}. Conventional neutrino oscillation searches 
in vacuo are insensitive to this mixing. However, in such neutrinos
may resonantly convert in the dense medium of a supernova
\cite{massless0,massless}. One can show how the observation of 
the energy spectrum of the SN1987A $\bar{\nu}_e$'s \cite{ssb} 
and the $r$-process nucleosynthesis in the supernova \cite{qian}
may be used to provide very stringent constraints on {\sl massless} 
neutrino mixing angles, as illustrated in \fig{SN87} and \fig{rprocess},
respectively.  The regions to the right of the solid curves 
are forbidden, those to the left are allowed. For detailed
explanation see ref. \cite{massless}.

As we saw above despite all the limits from laboratory
experiments, both at accelerators and reactors, as well
as the limits from cosmology and astrophysics, there is
considerable room for interesting new effects in the neutrino
sector. In addition to searches for neutrino-less double beta 
decay and neutrino oscillations, specially at long baseline 
experiments, it is worthwhile to continue the efforts to 
improve present laboratory  limits on \neu mass. One method 
sensitive to large masses is to search for distortions in the 
energy spectra of leptons coming from $\pi, K$ weak decays such 
as $\pi, K \ra e \nu$, $\pi, K \ra \mu \nu$, as well as 
kinks in nuclear $\beta$ decays.

\section{HINTS FOR NEUTRINO MASSES}
\vskip .3cm

Detecting nonzero neutrino masses could be very far reaching 
for the understanding of fundamental issues in particle physics, 
astrophysics, as well as the large scale structure of our universe.
So far the only indications in favour of nonzero neutrino rest
masses have been provided by astrophysical and cosmological 
observations, with a varying degree of theoretical input.  
We now turn to these.

\subsection{Dark Matter}
\vskip .3cm

Considerations based on structure formation in the Universe have
become a popular way to argue in favour of the need of a massive
neutrino \cite{cobe2}. Indeed,  by combining the observations of 
cosmic background temperature 
anisotropies on large scales performed by the COBE  satellite 
\cite{cobe} with cluster-cluster correlation data e.g. from 
IRAS \cite{iras} one finds that it is not possible to fit
well the data on all scales within the framework of the
simplest cold dark matter (CDM) model. The simplest way to
obtain a good fit is to postulate that there is a mixture 
of cold and hot components, consisting of about 70 \% CDM 
with about 25 \% {\sl hot dark matter} (HDM) and a small 
amount in baryons \cite{cobe2}.
The best candidate for the hot dark matter component 
is  a massive neutrino in the few eV mass range.
It has been argued that this could be the tau neutrino,
in which case one might expect the existence of \ne $\ra$ \nt 
or \nm $\ra$ \nt oscillations. Searches for these oscillations
are now underway at CERN, with a similar proposal also at 
Fermilab \cite{chorus}. This mass scale is also consistent 
with the hints in favour of neutrino oscillations reported 
by the LSND experiment \cite{lsnd}.

\subsection{Solar Neutrinos}
\vskip .3cm

So far the averaged data collected by the chlorine \cite{cl}, 
Kamiokande \cite{k}, as well as by the low-energy data on pp 
neutrinos from the GALLEX and SAGE experiments \cite{ga,sa} 
still pose a persisting puzzle. The most recent data can be
summarised as:
\bea
\label{data}
R_{Cl}^{exp}= (2.55 \pm 0.25) \mbox{SNU} \\ \nonumber
R_{Ga}^{exp}= (74 \pm 8) \mbox{SNU}  \\ \nonumber
R_{Ka}^{exp}= (0.44 \pm 0.06) R_{Ka}^{BP95} 
\eea 
where  $R_{Ka}^{BP95}$ is the BP95 standard solar model (SSM) 
prediction of ref. \cite{SSM}.
For the gallium result we have taken the average of the GALLEX \cite{ga} 
and the SAGE measurements \cite{sa}. 

Comparing the data of gallium experiments with the Kamiokande data
one sees the need for a reduction of the $^7$Be flux relative to 
standard solar model \cite{SSM} expectations. Inclusion of the 
Homestake data only sharpens the discrepancy, suggesting that the 
solar \neu problem is indeed a real problem. The totality of the 
data strongly suggests that the simplest astrophysical solutions 
are ruled out, and that new physics is needed \cite{CF}. The most 
attractive possibility is to assume the existence of \neu 
conversions involving very small \neu masses. In the framework 
of the MSW effect \cite{MSW} the required solar neutrino 
parameters $\Delta m^2$ and $\sin^2 2\theta$ are determined 
through a $\chi^2$ fit of the experimental data
\footnote{For simplicity we neglect theoretical uncertainties,
earth effects, as well as details of the neutrino production 
region.}. 
Fig. 3, taken from ref. \cite{noise}, shows the 90\% C.L.
areas for the in the BP95 model for the case of active neutrino 
conversions. The fit favours the small mixing solution over the 
large mixing one, due mostly to the larger reduction of the 
$^7$Be flux found in the former.
\begin{figure}
\psfig{file=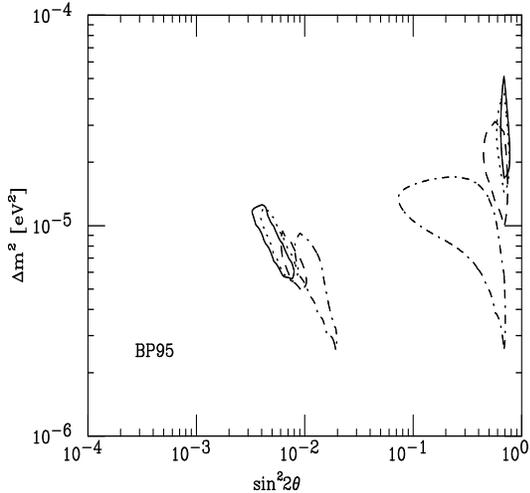,width=8.5cm,height=8cm,angle=90}
\vglue -8.1cm
\hglue 2.5cm
\psfig{file=cover2.ps,height=0.7cm,width=3cm,angle=90}
\vglue 7.2cm
\vglue -1.3cm
\caption{Allowed solar neutrino oscillation parameters for
active neutrino conversions.}
\vglue -0.5cm
\label{msw}
\end{figure}
Here the different regions in dashed and dot-dashed are
associated with a non-zero level of random fluctuations
fluctuations in the solar matter density \cite{BalantekinLoreti}.
The existence of such noise-type fluctuations is not excluded by 
the SSM nor by present helioseismology studies. The solid curves 
are for the standard case of zero matter density random fluctuations, 
$\xi = \delta \rho / \rho = 0$, corresponding to a smooth solar 
density profile. The regions inside the other curves correspond 
to the case where matter density fluctuations are assumed. Noise 
causes a slight shift of $\Delta m^2$ towards lower values and a 
larger shift of $\sin ^2 2 \theta$ towards larger values. 
The corresponding allowed $\Delta m^2$ range for $\xi = 8$ \% is 
$2.5 \times 10^{-6} <\Delta m^2< 9 \times 10^{-6}$ eV$^2$
instead of 
$5 \times 10^{-6} <\Delta m^2< 1.2 \times 10^{-5}$ eV$^2$
in the noiseless case.
The large mixing area is less stable, with a tendency to shift 
towards smaller $\Delta m^2$ and $\sin^2 2 \theta$ values.

It is interesting to note that the $^7$Be neutrinos are the 
solar neutrino spectrum component which is most affected by 
the matter noise. Therefore the Borexino experiment should 
be an ideal tool for studying the solar matter fluctuations, 
if sufficiently small errors can be achieved. Its potential in 
"testing" the level of solar matter density fluctuations is 
discussed in ref. \cite{noise}, which also contains a 
discussion of sterile solar neutrino conversions, as well 
as a comparison with other solar models.

\subsection{Atmospheric Neutrinos}
\vskip .3cm

The Kamiokande and IMB underground experiments, and possibly 
also Soudan2, have indications which support an apparent deficit 
in the expected flux of atmospheric $\nu_\mu$'s relative to that 
of $\nu_e$'s that would be produced from conventional decays of 
$\pi$'s, $K$'s as well as secondary muon decays \cite{Barish}. 
Although the predicted absolute fluxes of \neus produced by 
cosmic-ray interactions in the atmosphere are uncertain at the 
20\% level, their ratios are expected to be accurate to within 
5\%. While some of the experiments, such as Frejus and NUSEX, 
have not found a firm evidence, it has been argued that there 
may be a strong hint for an atmospheric neutrino deficit that 
could be ascribed to \neu oscillations.
Recent results from Kamiokande on higher energy \neus 
strengthen the case for an atmospheric \neu problem. The
relevant oscillation parameters were discussed here
by Inoue \cite{atm}.

\section{RECONCILING PRESENT HINTS}
\vskip .3cm

\subsection{Almost Degenerate Neutrinos}
\vskip .3cm

It is not easy to account for the three observations cosmology 
and astrophysics discussed above in a framework containing just 
the three known \neus. The only possibility to fit these 
observations in a three-neutrino world is if all of them have 
nearly the same mass $\sim$ 2 eV \cite{caldwell}. This can be 
arranged, for example in general seesaw models which also contain 
an effective triplet vacuum expectation value \cite{2227,LR} 
contributing to the light neutrino masses. This term should 
be added to \eq{SEESAW}. Thus one can construct extended 
seesaw models where the main contribution to the light 
\neu masses ($\sim$ 2 eV) is universal, due to a suitable 
horizontal symmetry, while the splittings between \ne and 
\nm explain the solar \neu deficit and that between \nm 
and \nt explain the atmospheric \neu anomaly \cite{DEG}.
For a study of the required parameters, see ref. \cite{mina}.

\subsection{Four-Neutrino Models}
\vskip .3cm

The simplest way to fit all the data is to add a 
fourth \neu species which, from the LEP data on the 
invisible Z width, we know must be of the sterile type,
call it \ns. The first scheme of this type gives mass
to only one of the three neutrinos at the tree level,
keeping the other two massless \cite{OLD}. 
In a seesaw scheme with broken lepton number, radiative 
corrections involving gauge boson exchanges will give 
small masses to the other two neutrinos \ne and \nm
\cite{Choudhury}. However, since the singlet \neu is 
super-heavy in this case, there is no room to account 
for the three hints discussed above.

Two basic schemes have been suggested to keep the sterile
neutrino light due to a special symmetry. In addition to the
sterile \neu \ns, they invoke additional Higgs bosons beyond 
that of the Standard Model, in order to generate radiatively
the scales required for the solar and atmospheric \neu
conversions. In these models the \ns either lies at the dark matter 
scale \cite{DARK92} as illustrated in \fig{pv}
\begin{figure}
\psfig{file=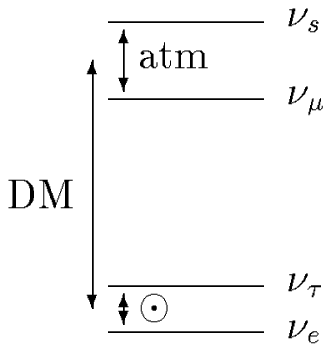,width=4.5cm,height=5cm}
\caption{"Heavy" Sterile 4-Neutrino Model}
\label{ptv}
\vglue -1cm
\end{figure}
or, alternatively, at the solar \neu scale \cite{DARK92B}. 
\begin{figure}
\psfig{file=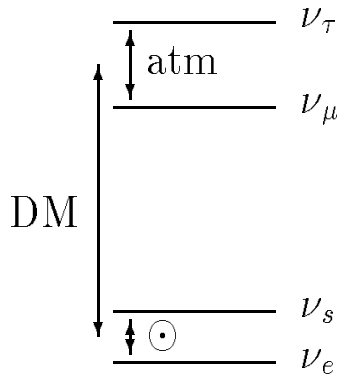,width=4.5cm,height=5cm}
\caption{Light Sterile 4-Neutrino Model}
\label{pv}
\vglue -.8cm
\end{figure}
In the first case the atmospheric
\neu puzzle is explained by \nm to \ns oscillations,
while in the second it is explained by \nm to \nt
oscillations. Correspondingly, the deficit of
solar \neus is explained in the first case
by \ne to \nt oscillations, while in the second 
it is explained by \ne to \ns oscillations. In both 
cases it is possible to fit all observations together. 
However, in the first case there is a clash with the 
bounds from big-bang nucleosynthesis. In the latter 
case the \ns is at the MSW scale so that nucleosynthesis
limits are satisfied. They nicely agree with the best 
fit points of the atmospheric neutrino parameters
from Kamiokande \cite{atm}. Moreover, it can naturally fit the 
hints of neutrino oscillations of the LSND experiment \cite{lsnd}.
For a more general study of the required parameters in
four-neutrino models, see ref. \cite{okada}.

Another theoretical possibility is that all active
\neus are very light, while the sterile \neu \ns is
the single \neu responsible for the dark matter
\cite{DARK92D}.

\subsection{Mev Tau Neutrino}
\vskip .3cm

An MeV range tau neutrino is an interesting possibility 
to consider for two reasons. First, such mass is within 
the range of the detectability, for example at a tau-charm
factory \cite{jj}. On the other hand, if such neutrino 
decays  before the matter dominance epoch, its decay
products would add energy to the radiation, thereby 
delaying the time at which the matter and radiation 
contributions to the energy density of the universe 
become equal. Such delay would allow one to reduce
the density fluctuations on the smaller scales purely 
within the standard cold dark matter scenario, and could 
thus reconcile the large scale fluctuations observed by
COBE \cite{cobe} with the observations such as those 
of IRAS \cite{iras} on the fluctuations on smaller scales.

In ref. \cite{JV95} a model was presented where an unstable 
MeV Majorana tau \neu naturally reconciles the cosmological 
observations of large and small-scale density fluctuations 
with the cold dark matter model (CDM) and, simultaneously, 
with the data on solar and atmospheric neutrinos discussed
above. The solar \neu deficit is explained through long 
wavelength, so-called {\sl just-so} oscillations \cite{just-so}
involving conversions of \ne into both \nm and a sterile species \ns, 
while the atmospheric \neu data are explained through \nm 
$\ra$ \ne conversions. Future long baseline \neu oscillation 
experiments, as well as some reactor experiments will test 
this hypothesis. The model assumes the spontaneous violation
of a global lepton number symmetry at the weak scale. 
The breaking of this symmetry generates the cosmologically 
required decay of the \nt with lifetime
$\tau_{\nu_\tau} \sim 10^2 - 10^4$ seconds, as well 
as the masses and oscillations of the three light 
\neus \ne, \nm and \ns required in order to account for 
the solar and atmospheric \neu data. One can verify 
that the big-bang nucleosynthesis constraints 
\cite{KTCS91,DI93} can be satisfied in this model. 

\section{CONCLUSION}
\vskip .3cm

Although theory alone can not predict neutrino masses, 
it is certainly true that neutrino masses are strongly suggested 
by present theoretical models of elementary particles. On the other 
hand, they seem to be required to account for present astrophysical 
and cosmological observations. Neutrino mass studies in nuclear 
$\beta$ decays and peak search experiments should continue. 
Searches for $\beta \beta_{0\nu}$ decays with enriched germanium 
could test the quasi-degenerate neutrino scenario of section 5.1. 
Underground experiments at Superkamiokande, Borexino, and Sudbury 
will shed more light on the solar neutrino issue. 
Oscillation searches in the \ne $\ra$ \nt and \nm $\ra$ \nt 
channels at accelerators should soon improve over the present 
situation illustrated in \fig{oscil}, while long-baseline experiments 
both at reactors and accelerators are being considered. The latter will 
test the regions of oscillation parameters presently suggested by 
atmospheric \neu anomaly. Finally, new satellite experiments 
will test different models of structure formation, and shed 
light on the possible role of neutrinos as dark matter. 
If neutrinos are massive they could be responsible for a 
wide variety of implications, covering an impressive range 
of energies. These could be probed in experiments performed 
at underground installations as well as particle accelerators. 

\bibliographystyle{ansrt}

\end{document}